\documentclass[prd,preprint,nofootinbib,superscriptaddress]{revtex4}
\usepackage[latin9]{inputenc}
\usepackage{amsmath}
\usepackage{amssymb}
\usepackage{graphicx}

\makeatletter

\providecommand{\tabularnewline}{\\}

\@ifundefined{textcolor}{}
{%
 \definecolor{BLACK}{gray}{0}
 \definecolor{WHITE}{gray}{1}
 \definecolor{RED}{rgb}{1,0,0}
 \definecolor{GREEN}{rgb}{0,1,0}
 \definecolor{BLUE}{rgb}{0,0,1}
 \definecolor{CYAN}{cmyk}{1,0,0,0}
 \definecolor{MAGENTA}{cmyk}{0,1,0,0}
 \definecolor{YELLOW}{cmyk}{0,0,1,0}
}

\makeatother

\begin{document}
\title{Comparing early dark energy and extra radiation solutions to the Hubble
tension with BBN }
\author{Osamu Seto}
\email{seto@particle.sci.hokudai.ac.jp}

\affiliation{Institute for the Advancement of Higher Education, Hokkaido University,
Sapporo 060-0817, Japan}
\affiliation{Department of Physics, Hokkaido University, Sapporo 060-0810, Japan}
\author{Yo Toda}
\email{y-toda@particle.sci.hokudai.ac.jp}

\affiliation{Department of Physics, Hokkaido University, Sapporo 060-0810, Japan}
\begin{abstract}
A shorter sound horizon scale at the recombination epoch, arising
from introducing extra energy components such as extra radiation or
early dark energy (EDE), is a simple approach resolving the so-called
Hubble tension. We compare EDE models, an extra radiation model, and
a model in which EDE and extra radiation coexist, paying attention
to the fit to big bang nucleosynthesis (BBN). We find that the fit
to BBN in EDE models is somewhat poorer than that in the $\Lambda$CDM
model, because the increased inferred baryon asymmetry leads to a
smaller deuterium abundance. We find that an extra radiation-EDE coexistence
model gives the largest present Hubble parameter $H_{0}$ among the
models studied. We also the examine data sets dependence, whether
we include BBN or not. The difference in the extra radiation model
is $3.22<N_{\mathrm{eff}}<3.49\,(68\%)$ for data sets without BBN
and $3.16<N_{\mathrm{eff}}<3.40\,(68\%)$ for data sets with BBN,
which is so large that the $1\sigma$ border of the larger side becomes
the $2\sigma$ border. 
\end{abstract}
\preprint{EPHOU-21-002}

\maketitle
\vspace*{1cm}

%\makeatletter

%%%%%%%%%%%%%%%%%%%%%%%%%%%%%% LyX specific LaTeX commands.
%% Because html converters don't know tabularnewline
%\providecommand{\tabularnewline}{\\}

%\makeatother

%\begin{document}

\section{Introduction}

%%%%%%%%%%%%%%%%%%%%%%%

The $\Lambda$CDM cosmological model has been successful in explaining
the properties and evolution of our Universe. However, various low-redshift
measurements of the Hubble constant $H_{0}$ have reported a significantly
larger value than that inferred from the temperature anisotropy of
cosmic microwave background (CMB) measured by \textsl{Planck} (2018)
$H_{0}=67.36\pm0.54$ km/s/Mpc~\cite{Aghanim:2018eyx}. SH0ES measures
the Hubble constant by using Cepheids and type Ia supernovae as standard
candles and has reported $H_{0}=73.45\pm1.66$ km/s/Mpc in Ref.~\cite{Riess:2018uxu}
(R18) and $H_{0}=74.03\pm1.42$ km/s/Mpc in Ref.~\cite{Riess:2019cxk}
(R19). Similarly, the H0LiCOW Collabration obtained the result $H_{0}=73.3\pm1.7$
km/s/Mpc from gravitational lensing with time delay~\cite{Wong:2019kwg}.
%Miras, MCP
Since all local measurements of $H_{0}$ by different methods consistently
indicate a larger value of $H_{0}$ than that from \textsl{Planck},
even if there is an unknown systematic error~\cite{Efstathiou:2013via,Freedman:2017yms,Rameez:2019wdt,Ivanov:2020mfr},
this discrepancy cannot be easily solved~\cite{Bernal:2016gxb}.

As this discrepancy seems serious, several ideas for an extension
of the $\Lambda$CDM model have been proposed to solve or relax this
tension. A modification in the early Universe would be more promising
than our during later times, because low-redshift $z$ cosmology is
also well constrained by baryon acoustic oscillation (BAO) measurements.
One approach to relax the Hubble tension is to introduce a beyond-the-standard-model
component~\cite{Bernal:2016gxb,Aylor:2018drw}. One of the simplest
methods is to increase the relativistic degrees of freedom parametrized
by $N_{\mathrm{eff}}$~\cite{Aghanim:2018eyx}. By quoting the value
\begin{equation}
N_{\mathrm{eff}}=3.27\pm0.15(68\%),\label{Eq:PlanckNeff}
\end{equation}
for (CMB$+$BAO$+$R18) from the \textsl{Planck} Collaboration~\cite{Aghanim:2018eyx},
it has been regarded that $0.2\lesssim\Delta N_{\mathrm{eff}}\lesssim0.5$
is preferred to relax the $H_{0}$ tension. Several beyond-the-standard-model
proposals~\cite{DEramo:2018vss,Escudero:2019gzq} could accommodate
such an extra $N_{\mathrm{eff}}$. Some of them are interesting because
they can address not only the Hubble tension, but also other subjects
such as the anomalous magnetic moment of the muon~\cite{Escudero:2019gzq},
the origin of neutrino masses \cite{Escudero:2019gvw} or a sub-GeV
weakly interacting massive particle dark matter~\cite{Okada:2019sbb}.
Another popular scenario is so-called early dark energy (EDE) models,
where a tentative dark energy component somewhat contributes the cosmic
expansion around the recombination epoch~\cite{Poulin:2018zxs,Poulin:2018dzj,Poulin:2018cxd,Agrawal:2019lmo,Alexander:2019rsc,Lin:2019qug,Smith:2019ihp,Berghaus:2019cls,Sakstein:2019fmf,Chudaykin:2020acu,Braglia:2020bym,Gonzalez:2020fdy,Niedermann:2020dwg,Lin:2020jcb,Murgia:2020ryi,Chudaykin:2020igl,Yin:2020dwl}.
The main idea of how to relax the Hubble tension by adding an extra
energy component is summarized as follows. %\begin{enumerate}
%\item

Once the cosmic expansion rate is enhanced by introducing a new extra
component, the comoving sound horizon for acoustic waves in a baryon-photon
fluid at the time of recombination with the redshift $z_{*}$ becomes
shorter than that in the standard $\Lambda$CDM model.

%\item 
The position of the first acoustic peak in the CMB temperature anisotropy
power spectrum corresponds to its angular size $\theta_{*}$, which
is related to the sound horizon $r_{s*}$ by $\theta_{*}\equiv r_{s*}/D_{M*}$,
with the angular diameter distance 
\begin{equation}
D_{M*}=\int_{0}^{z*}\frac{dz}{H(z)}.\label{Eq:DM*}
\end{equation}
For a fixed measured $\theta_{*}$, the reduction of $D_{M*}$ due
to the shorter $r_{s*}$ leads to a larger Hubble parameter, because
the primary term of the integrand in Eq.~(\ref{Eq:DM*}) at low $z$
does not change much.

%\item 
The effects of the shorter sound horizon due to new components like
those mentioned above can be seen in the power spectrum as a shift
of peak positions to higher multipoles $\ell$. Other effects are
to let the first and second peaks higher as well as other high $\ell$
peaks lower. The magnitudes of the change of the peak heights and
position shifts depend on the specific extra component model. On the
other hand, as is well known, generally an increase of the dark matter
density shifts the spectra to lower $\ell$ values and reduces the
height of the peaks, and an increase of the baryon density extends
the height difference between the first and second peaks~\cite{Hu:2001bc}.
The scalar spectral index $n_{s}$ controls the spectral tilt of the
whole range of the spectrum. In order to compensate the new component
effects on the power spectrum, the dark matter density needs to be
increased to return the original spectrum that matches with the $\Lambda$CDM,
and then the baryon density also needs to be increased to adjust the
relative height of the first and second peaks. %\end{enumerate}

This approach is limited due to the resultant modification to the
Silk damping scale, i.e., the photon diffusion scale~\cite{Knox:2019rjx}.
One way to see the first problem is the fact that, as mentioned above,
the phase shift and damping of the amplitude of high-$\ell$ peaks
cannot be well recovered by changing only the dark matter and baryon
density\footnote{For a scenario free from this diffusion problem, see Refs.~\cite{Sekiguchi:2020teg,Sekiguchi:2020igz}.}.
For EDE models, a poor fit to large-scale structure data has also
been claimed~\cite{Hill:2020osr,Ivanov:2020ril}, which can be seen
in the value of $\sigma_{8}$ in Ref.~\cite{Poulin:2018cxd}. Despite
this limitation, the introduction of extra radiation or EDE is a simple
extension of $\Lambda$CDM that addresses the $H_{0}$ discrepancy.
The extra radiation energy and EDE contribute differently throughout
the whole cosmological history. While EDE significantly contributes
to the energy budget only around the epoch of matter-radiation equality
to recombination and its energy density decreases quickly, the extra
radiation exists throughout the whole history of the Universe. This
difference can be seen in its effects on big bang nucleosystheisis
(BBN). In fact, a study for $N_{\mathrm{eff}}$ with referring BBN
in the context of the Hubble tension was done in Refs.~\cite{Cuceu:2019for,Schoneberg:2019wmt}.
In EDE models, although the negligible energy density of the EDE component
at the BBN epoch appears not to change the BBN prediction, the baryon
abundance inferred from the CMB would be different from that in the
$\Lambda$CDM model to adjust the CMB spectrum and hence the resultant
light element abundance could be affected. In this paper, by taking
the fit with BBN into account, we evaluate and compare these scenarios
of additional relativistic degrees of freedoms and EDE.

This paper is organized as follows. In the next section, we first
describe our modeling of the extra radiation and EDE. After we describe
the methods and data sets used in our analysis in Sec.~\ref{sec:analysis},
we show the results and discuss their interpretation in Sec.~\ref{sec:results}.
The last section is devoted to a summary.

%%%%%%%%%%%%%%%%%%%%%%%

\section{Modeling}

The expansion rate of the Universe---the Hubble parameter,---is
defined as % 
\begin{equation}
H(t)=\frac{\dot{a}}{a},\label{eq:Hubble_t}
\end{equation}
where $a(t)$ is the scale factor and a dot denotes a derivative with
respect to cosmic time $t$. In the following, we use the scale factor
$a$ instead of time $t$ as a ``time'' variable. Then, we regard
the Hubble parameter as a function of $a$, $H(a)$ and normalize
the scale factor as $a(t_{0})=a_{0}=1$, with $t_{0}$ being the age
of the Universe.

\subsection{Extra radiation}

One simple ``solution'' to the Hubble tension is to increase the
effective number of neutrinos $N_{\mathrm{eff}}$, which is expressed
as 
\begin{equation}
\Omega_{r}=\left(1+\frac{7}{8}\left(\frac{4}{11}\right)^{4/3}N_{\mathrm{eff}}\right)\Omega_{\gamma}.
\end{equation}
Here, 
\begin{equation}
\Omega_{i}=\left.\frac{\rho_{i}}{3M_{P}^{2}H^{2}}\right|_{t=t_{0}},
\end{equation}
with $M_{P}$ being the reduced Planck mass, are the present values
of the density parameters for $i$ species. $\gamma$ and $\nu$ stand
for CMB photons and neutrinos, respectively. In the following, we
call this an $N_{\mathrm{eff}}$ model.

\subsection{Early dark energy}

In the literature, the EDE scenarios are modeled by a variety of functional
forms for scalar fields, such as axion-potential-like~\cite{Poulin:2018zxs,Poulin:2018dzj,Poulin:2018cxd},
polynomial~\cite{Agrawal:2019lmo}, acoustic dark energy~\cite{Lin:2019qug},
$\alpha$-attractor-like~\cite{Braglia:2020bym}, and others. Those
properties and differences due to various potential forms have been
studied in those literatures. In contrast, in this work, since our
main focus is on the differences between the EDE and the extra radiation,
we adopt a simple fluid picture. In order to treat time-dependent
dark energy, we define the dark energy equation of state $w(a)=p_{DE}(a)/\rho_{DE}(a)$.
By integrating the continuity equation 
\begin{equation}
\dot{\rho}_{DE}+3H\rho_{DE}(1+w(a))=0,
\end{equation}
we obtain~\cite{Copeland:2006wr} 
\begin{align}
\rho_{DE}(a)=\rho_{DE}(a_{0})\exp\left(-\int_{a}^{a_{0}}3(1+w(\hat{a}))\frac{d\hat{a}}{\hat{a}}\right).\label{Eq:rho_DE}
\end{align}
We find the cosmic expansion by solving the Friedman equation for
the EDE models 
\begin{align}
\frac{H^{2}(a)}{H_{0}^{2}} & =(\Omega_{c}+\Omega_{b})a^{-3}+\Omega_{r}a^{-4}+\Omega_{DE}(a),\label{Eq:Freidman_EDE}\\
\Omega_{DE}(a) & =\Omega_{DE}(a_{0})\exp\left(-\int_{a}^{a_{0}}3(1+w(\hat{a}))\frac{d\hat{a}}{\hat{a}}\right).\label{Eq:Omega_DE}
\end{align}
The indices of each $\Omega$ parameter, $c,b,r,~\mathrm{and}~DE$,
stand for cold dark matter, baryons, radiation, and dark energy, respectively.
Here, DE represents the sum of the EDE component and the cosmological
constant $\Lambda$ which is responsible for the present accelerating
cosmic expansion as $\rho_{\mathrm{DE}}(a)=\rho_{\mathrm{EDE}}(a)+\rho_{\Lambda}$.

\begin{figure}[htbp]
\centering \includegraphics[width=15cm]{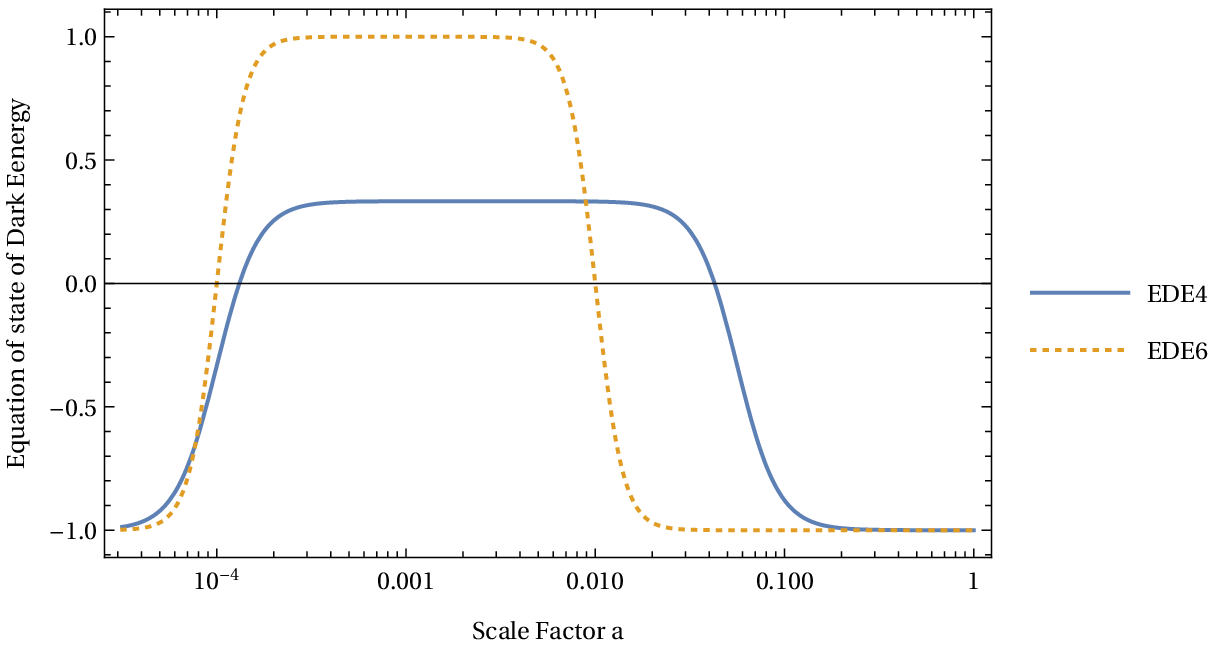} \includegraphics[width=15cm]{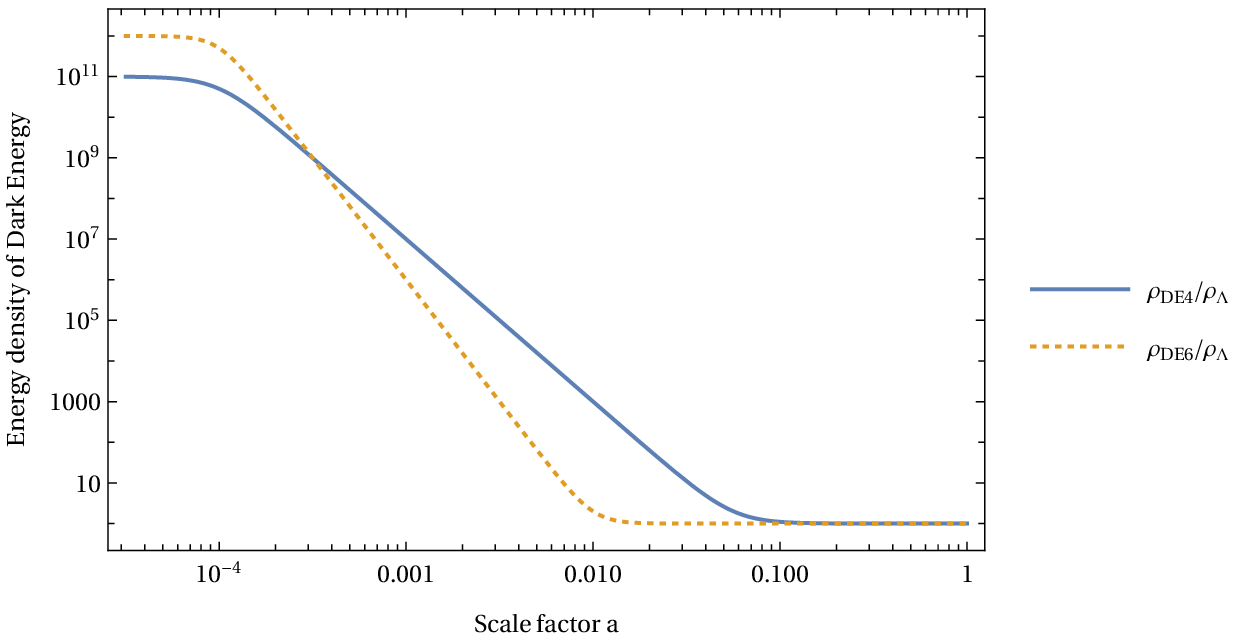}

\caption{Evolution of the equation-of-state parameter $w_{n}$ and $\rho_{DEn}(a)/\rho_{\Lambda}$
for $n=4$ and $6$. For definiteness, in these plots we take $a_{c}=10^{-4},a_{m}=10^{-5/4}$
for $n=4$ and $a_{c}=10^{-4},a_{m}=10^{-2}$ for $n=6$. We can confirm
that $\rho$ and $w$ behave like radiation or kination for $a_{c}<a<a_{m}$
and like a cosmological constant for $a<a_{c}$ and $a_{m}<a$.}
\label{Fig:EDE:evolution} 
\end{figure}

\begin{figure}[htbp]
\includegraphics[width=17cm]{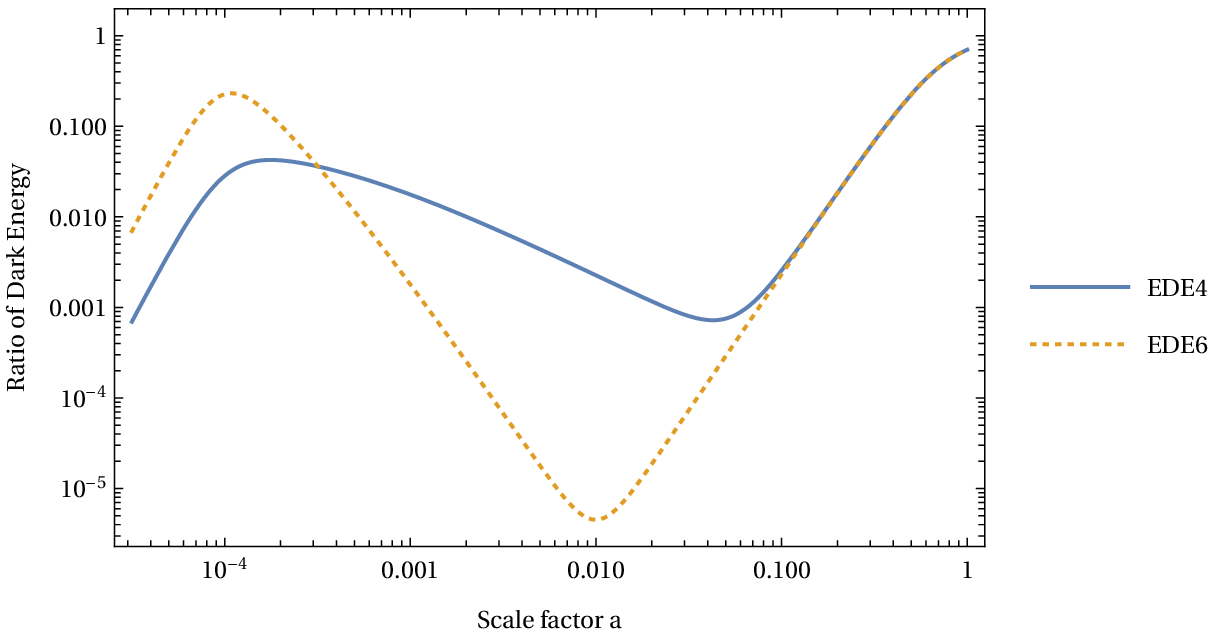} \caption{Evolution of $f_{\mathrm{EDE}}(a)$. The choices of $a_{c}$ and $a_{m}$
are the same as in Fig.~\ref{Fig:EDE:evolution}.}
\label{Fig:EDE:evolution_in_f} 
\end{figure}

In the EDE scenarios, the energy component of dark energy becomes
significant around the moment of the matter-radiation equality and
contributes to about several percents of the total energy density.
Soon after that moment, the EDE component decreases as $\rho_{\mathrm{DE}}\propto a^{-n}$
and faster than the background energy densities do. Here we introduce
two parameters: $a_{c}$ and $a_{m}$. $a_{c}$ is the scale factor
at the moment when the EDE starts to decrease like radiation $(n=4)$
or kination $(n=6)$. $a_{m}$ is the scale factor at the moment when
the EDE component becomes as small as the cosmological constant. Phenomenologically,
we parametrize the equation of state $w(a)$ as 
\begin{equation}
w_{n}(a)=-1+\frac{n}{3}\frac{(a_{c}^{n}+a_{m}^{n})a^{n}}{(a_{c}^{n}+a^{n})(2a_{c}^{n}+a_{m}^{n}+a^{n})},\label{Eq:w}
\end{equation}
and the resultant $\Omega$ parameter is given by 
\begin{equation}
\Omega_{\mathrm{DE}n}(a)=\Omega_{\Lambda}\frac{a^{n}+2a_{c}^{n}+a_{m}^{n}}{a^{n}+a_{c}^{n}}.\label{Eq:OmegaEDE}
\end{equation}
For $a<a_{c}$ and $a_{m}<a$, the dark energy behaves like a cosmological
constant. Typical evolutions of $w_{n}(a)$ and $\rho_{DEn}(a)$ normalized
by $\rho_{\Lambda}$ are shown in Fig.~\ref{Fig:EDE:evolution}.
We note that the above $\Omega_{\mathrm{DE}n}(a)$ is proportional
to the $f_{\mathrm{EDE}}$ parameter often used in literature as 
\begin{equation}
f_{\mathrm{EDE}}(a)=\Omega_{\mathrm{DE}n}(a)\frac{H_{0}^{2}}{H^{2}(a)}.
\end{equation}
Typical evolutions of $f_{\mathrm{EDE}}(a)$ are shown in Fig.~\ref{Fig:EDE:evolution_in_f}.

Since Eq.~(\ref{Eq:OmegaEDE}) can be approximated as 
\begin{align}
\left.\Omega_{DEn}(a)\right|_{a=1}\simeq\Omega_{\Lambda}(1+a_{m}^{n})=\Omega_{\Lambda}+\Omega_{\textrm{EDE}},\label{eq:9}
\end{align}
we will use $\Omega_{\textrm{EDE}}/\Omega_{\Lambda}$ as a variable
parameter instead of $a_{m}$.

%%%%
%\centering \includegraphics[draft,width=15cm,type=eps]{graph001}
%\includegraphics[draft,width=15cm,type=eps]{graph002} 

Dealing perturbation, in this work we set its effective sound speed
$c_{s}^{2}=1$ for perturbations of the EDE component, motivated by
a class of scalar field models. One may compare this to each specific
scalar potential model in e.g., Refs.~\cite{Poulin:2018zxs,Poulin:2018dzj,Poulin:2018cxd,Alexander:2019rsc,Braglia:2020bym,Murgia:2020ryi,Chudaykin:2020acu,Chudaykin:2020igl}.
Our modeling will be closer to nonoscillatory scalar field models~\cite{Lin:2019qug,Braglia:2020bym}.

\section{Data and Analysis}

\label{sec:analysis}

We perform a Markov Chain Monte Carlo (MCMC) analysis on an $N_{\mathrm{eff}}$
model and the EDE model described in the previous section. We use
the public MCMC code \texttt{CosmoMC-planck2018}~\cite{Lewis:2002ah}
and implement the above EDE scenarios by modifying its equation file
in \texttt{CAMB}. For estimation of light elements, we use \texttt{PArthENoPE standard}~\cite{Pisanti:2007hk} in \texttt{CosmoMC}.

\subsection{Data sets}

We analyze models using the following cosmological observation data
sets. We include both temperature and polarization likelihoods for
high $l$ ($l=30$ to $2508$ in TT and $l=30$ to $1997$ in EE and
TE) and low$l$ \texttt{Commander} and lowE \texttt{SimAll} ($l=2$
to $29$) of Planck (2018) measurement of the CMB temperature anisotropy~\cite{Aghanim:2018eyx}.
We also include \textsl{Planck} lensing data~\cite{Aghanim:2018oex}.
For constraints on low-redshift cosmology, we include BAO data from
6dF~\cite{Beutler:2011hx}, DR7~\cite{Ross:2014qpa}, and DR12~\cite{Alam:2016hwk}.
We also include Pantheon data~\cite{Scolnic:2017caz} on the local
measurement of light curves and luminosity distance of supernovae,
as well as SH0ES (R19) data~\cite{Riess:2019cxk} on the local measurement
of the Hubble constant from the Hubble Space Telescope's observation
of Supernovae and Cephied variables. Finally, we include the data
sets on the helium mass fraction $Y_{P}$~\cite{Aver:2015iza} and
deuterium abundance D/H~\cite{Cooke:2017cwo} to impose the constraints
from BBN.

\subsection{EDE and neutrino parameter sets}

We take a prior range of $N_{\textrm{eff}}\in[2.2,3.6]$. This is
motivated by the $2\sigma$ limit $2.2\lesssim N_{\textrm{eff}}\lesssim3.6$
(BBN+$Yp$+D/H,$2\sigma$) in Ref.~\cite{Cyburt:2015mya}.

For the EDE model with $n=4$, which is denoted as EDE$4$ hereafter,
we fix $a_{c}=10^{-4}$ and vary parameters in the range $\Omega_{\textrm{EDE}}/\Omega_{\Lambda}\in[1\times10^{-7},1\times10^{-5}]$
. For the EDE model with $n=6$, which is denoted as EDE$6$ hereafter,
we fix $a_{c}=3\times10^{-4}$ and vary parameters in the range $\Omega_{\textrm{EDE}}/\Omega_{\Lambda}\in[1\times10^{-14},5\times10^{-12}]$.
These values of $a_{c}$ are motivated by the results in Ref.~\cite{Poulin:2018cxd}.

\section{Result and discussion}

\label{sec:results}

\subsection{Result}

\begin{figure}[htbp]
\includegraphics[width=12cm]{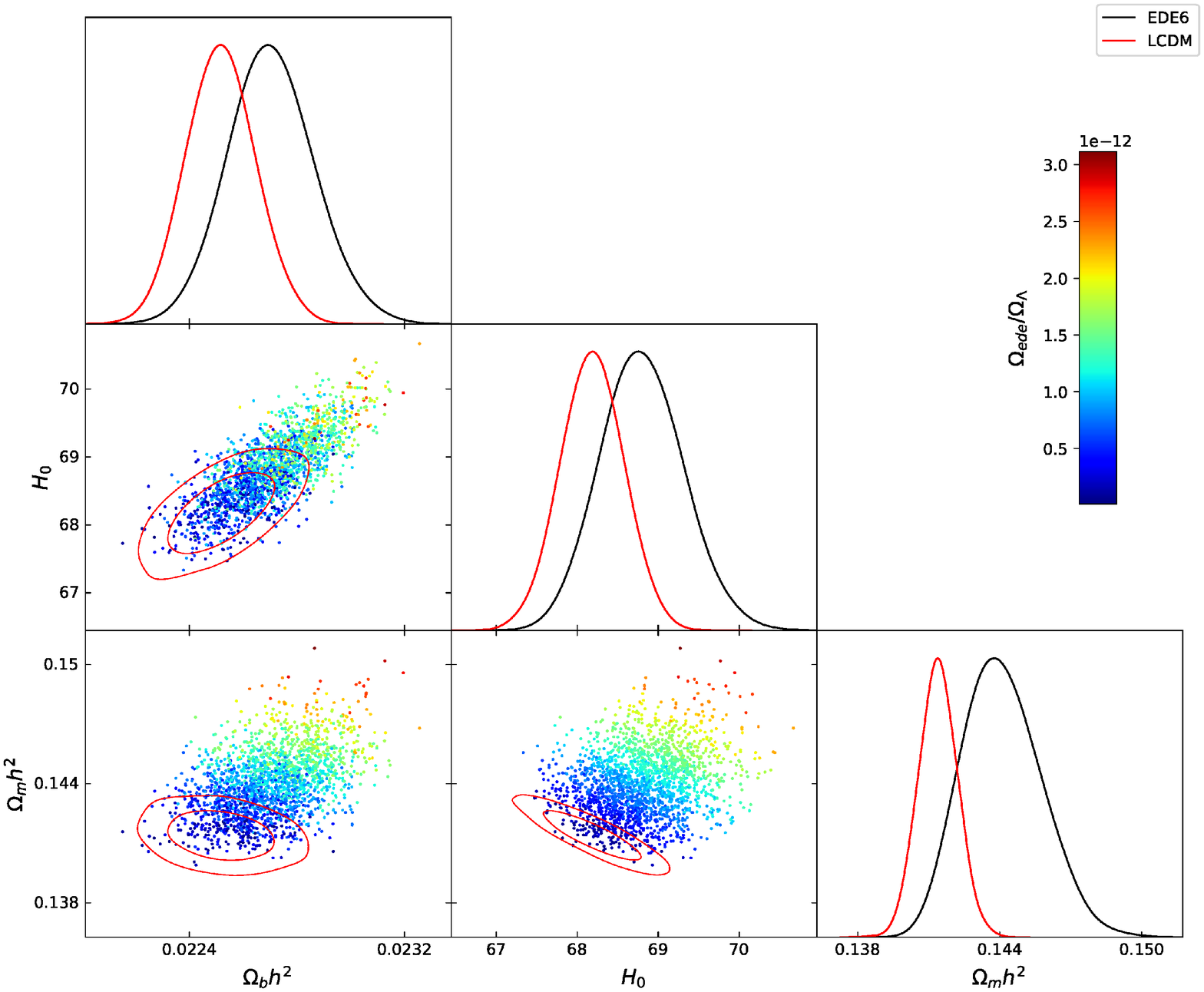} \caption{Posterior and constraints on the EDE$6$ model. Red curves are for
the $\Lambda$CDM model, for reference.}
\label{Fig:Kination-like} %\end{figure}
%\begin{figure}[hbtp]
\includegraphics[width=12cm]{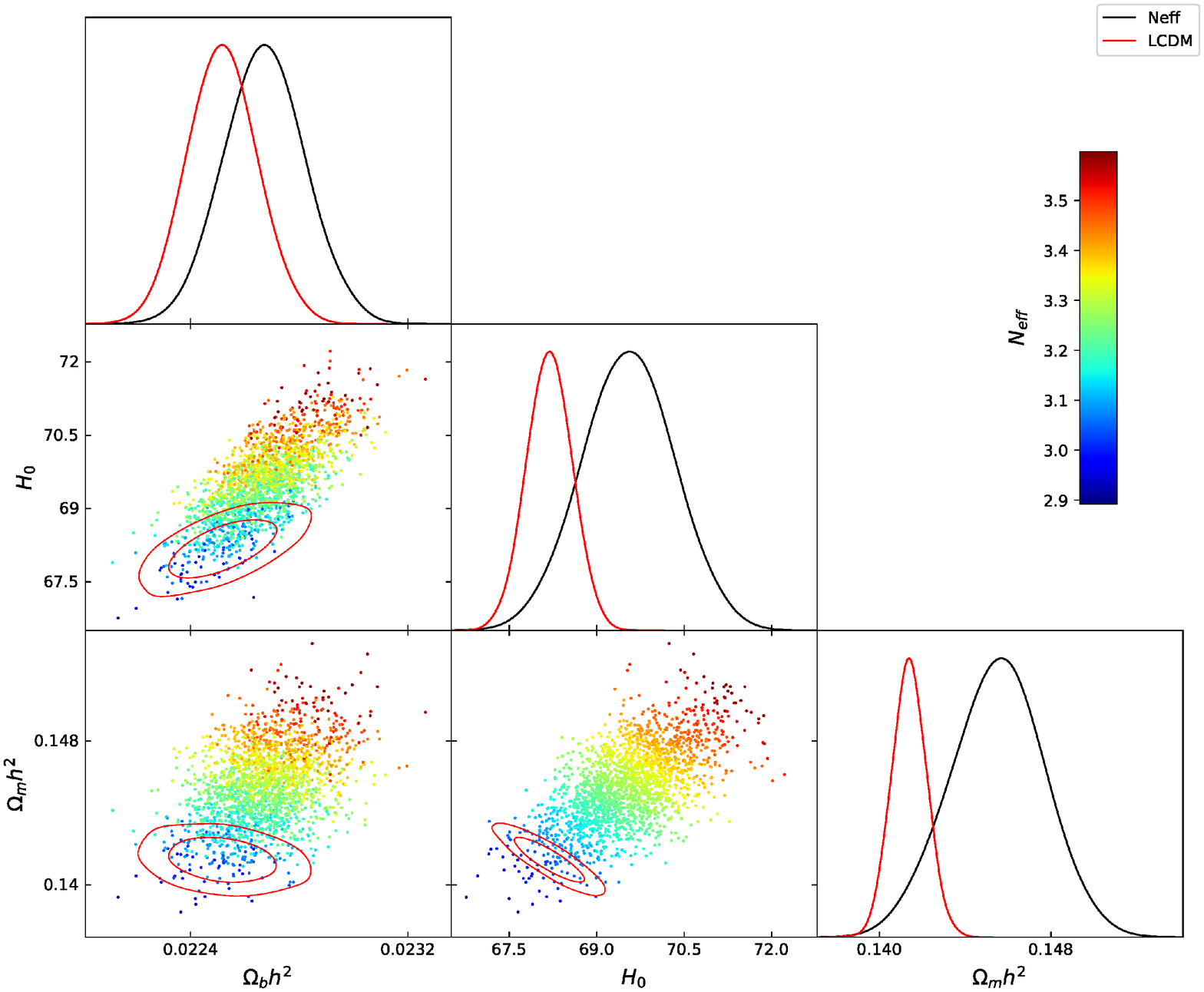} \caption{Posterior and constraints on the $N_{\mathrm{eff}}$ model. Red curves
are for the $\Lambda$CDM model, for reference.}
\label{Fig:Extra-Nu} 
\end{figure}

Although we have examined both the EDE$4$ and EDE$6$ models, we
have confirmed that the EDE$6$ model gives a slightly better fit
than the EDE$4$ model, as has been pointed out in previous works.
Thus, we show a posterior distribution for only the EDE$6$ model
in Fig.~\ref{Fig:Kination-like} and the posterior distribution for
an $N_{\mathrm{eff}}$ model in Fig.~\ref{Fig:Extra-Nu}. In addition
to the above two models which have been studied in literature, we
also consider the model where both the extra radiation and EDE components
exist, which hereafter we call the coexisting model or EDE$6+N_{\mathrm{eff}}$,
motivated by the fact that these are in principle independent sectors.

\begin{figure}[htbp]
\includegraphics[width=13cm]{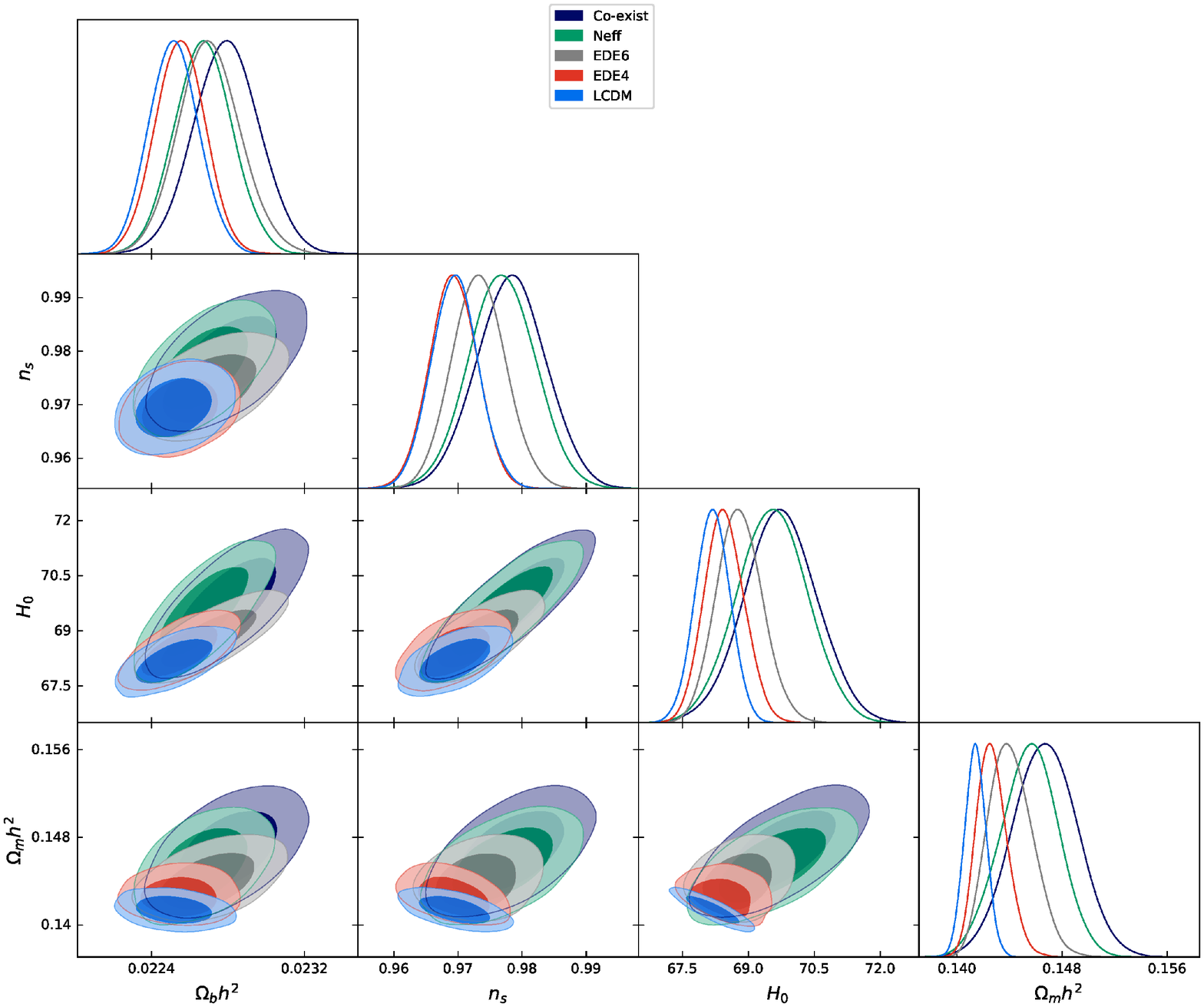} \caption{Posterior and constraints of several models.}
\label{Fig:Comparison} %\end{figure}
%\begin{figure}[htbp]
\includegraphics[width=13cm]{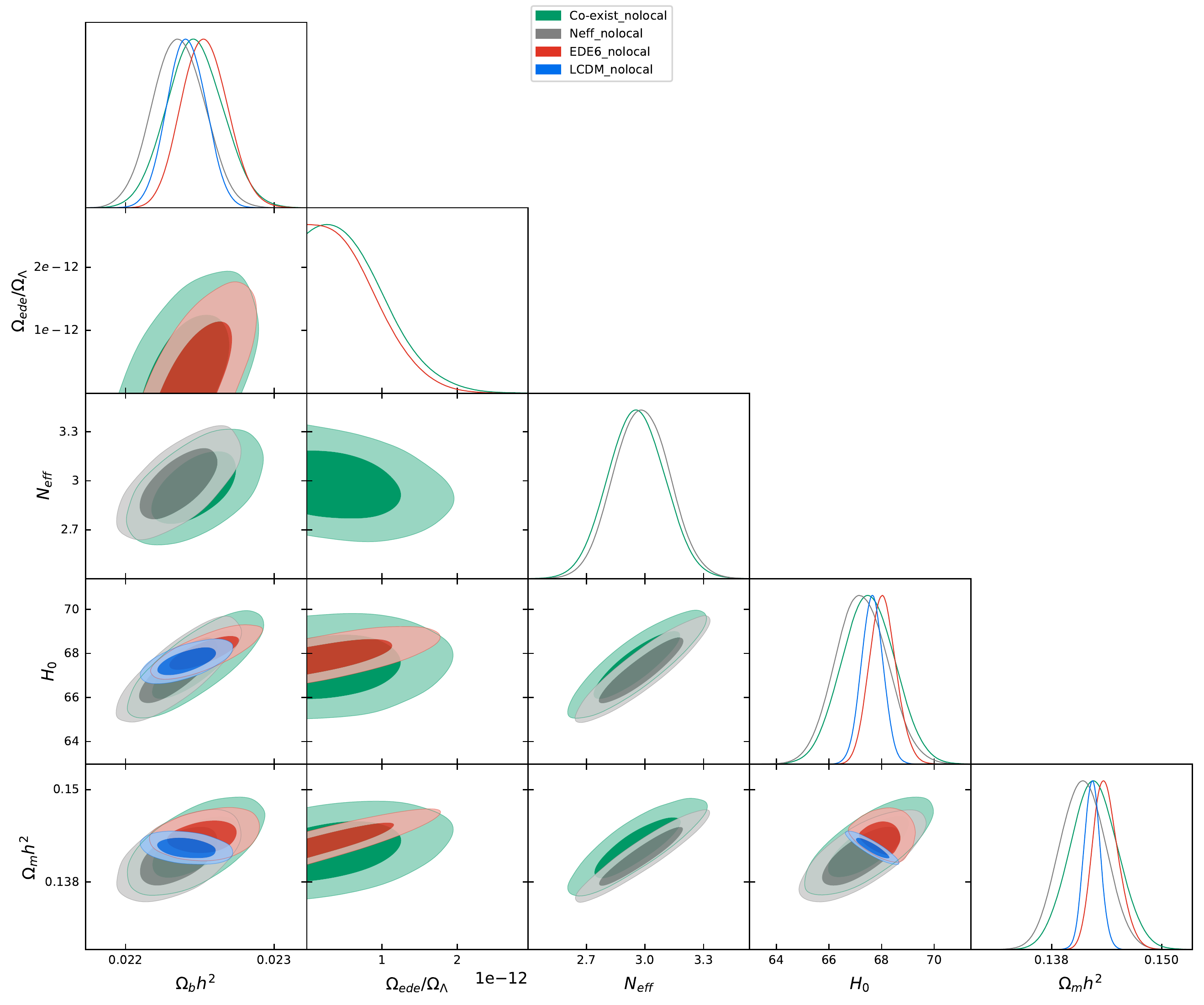} \caption{Posterior and constraints of several models without including R19
data.}
\label{Fig:nolocal} 
\end{figure}

To compare models, we show the combined plots of the EDE$4$, the
EDE$6$, the coexistence model with EDE$6$ plus $N_{\mathrm{eff}}$,
and $N_{\mathrm{eff}}$ together with $\Lambda$CDM for reference
in Fig.~\ref{Fig:Comparison}. These results are also summarized
in Table~\ref{Tab:Bestfit}. It is clear that both the EDE$6$ and
coexistence models prefer a larger value of $H_{0}$ than other models.
We confirm that the EDE$4$ model shows the poorest improvement for
the $H_{0}$ tension, as shown in Fig.~\ref{Fig:Comparison}.

For reference, we also show the same plot of the analysis without
including Pantheon and R19 data {[}in other words (CMB$+$BAO$+$BBN){]}
in Fig.~\ref{Fig:nolocal}, because one may wonder used data sets
dependence. One can confirm that the results for $N_{\mathrm{eff}}$
in Fig.~\ref{Fig:nolocal} almost reproduce Fig.~35 in Ref.~\cite{Aghanim:2018eyx}.
The constraints on EDE in Figs.~\ref{Fig:Comparison} and ~\ref{Fig:nolocal}
do not differ significantly. On the other hand, we can find sizable
shifts of the posterior and the cental values for $N_{\mathrm{eff}}$
and the coexintence model, depending on whether we include R19 data.
If one pays attention to only the central values, it looks like the
EDE indicates the largest value of $H_{0}$ and models with $N_{\mathrm{eff}}$
indicate even smaller values than $\Lambda$CDM does. However, the
constraints of a certain confidence level on the $N_{\mathrm{eff}}$
and the coexistence models are much weaker than those for the EDE
model. In addition, one should be aware of the presence of the prior
dependence in Fig.~\ref{Fig:nolocal}, where the energy density of
EDE is positive definite, while we allow $N_{\mathrm{eff}}<3.046$
for the models with $N_{\mathrm{eff}}$.

\begin{table}[htbp]
\begin{tabular}{|c||c|c|c|c|}
\hline 
Model  & EDE$6$  & EDE$6$+$N_{\mathrm{eff}}$  & $\Lambda$CDM  & $N_{\mathrm{eff}}$ \tabularnewline
\hline 
\hline 
$\Omega_{\textrm{EDE}}/\Omega_{\Lambda}$  & $\begin{array}{c}
1.00_{-0.65}^{+0.45}\times10^{-12}\\
(8.00\times10^{-13})
\end{array}$  & $\begin{array}{c}
0.79_{-0.71}^{+0.27}\times10^{-12}\\
(5.80\times10^{-13})
\end{array}$  & $\Lambda$CDM  & $\Lambda$CDM \tabularnewline
\hline 
$N_{\textrm{eff}}$  & 3.046  & $\begin{array}{c}
3.23\pm0.13\\
(3.18946)
\end{array}$  & 3.046  & $\begin{array}{c}
3.28\pm0.12\\
(3.28)
\end{array}$ \tabularnewline
\hline 
$H_{0}$ {[}km/s/Mpc{]}  & $\begin{array}{c}
68.80\pm0.52\\
(68.50)
\end{array}$  & $\begin{array}{c}
69.73\pm0.82\\
(69.42)
\end{array}$  & $\begin{array}{c}
68.19\pm0.39\\
(68.22)
\end{array}$  & $\begin{array}{c}
69.54\pm0.80\\
(69.38)
\end{array}$ \tabularnewline
\hline 
\hline 
%$n_{s}$  & $\begin{array}{c}
%0.9733\pm 0.0042\\
%(0.9697)
%\end{array}$  & $\begin{array}{c}
%0.9783\pm 0.0054\\
%(0.9772)
%\end{array}$ & $\begin{array}{c}
%0.9695\pm 0.0036\\
%(0.9701)
%\end{array}$  & $\begin{array}{c}
%0.9769\pm 0.0052\\
%(0.9770)
%\end{array}$ \\
%\hline 
CMB:lensing $\chi^{2}$  & 8.83  & 8.81  & 8.52  & 9.17 \tabularnewline
\hline 
CMB:TTTEEE $\chi^{2}$  & 2350.05  & 2352  & 2348.86  & 2354.72 \tabularnewline
\hline 
CMB:low$l$ $\chi^{2}$  & 22.04  & 21.48  & 22.83  & 21.93 \tabularnewline
\hline 
CMB:lowE $\chi^{2}$  & 395.88  & 397.82  & 399.52  & 396.51 \tabularnewline
\hline 
Cooke $\chi^{2}$  & 0.65  & 0.18  & 0.30  & 0.0037 \tabularnewline
\hline 
Aver $\chi^{2}$  & 0.23  & 0.92  & 0.22  & 1.56 \tabularnewline
\hline 
SH0ES $\chi^{2}$  & 15.18  & 10.56  & 16.75  & 10.71 \tabularnewline
\hline 
JLA Pantheon18 $\chi^{2}$  & 1034.85  & 1034.74  & 1034.77  & 1034.77 \tabularnewline
\hline 
BAO $\chi^{2}$  & 5.35  & 5.41  & 5.24  & 5.32 \tabularnewline
\hline 
%\hline 
%BBN $\chi^{2}$ & 1.29 & 1.82 & 0.67 & 1.90  \\
prior $\chi^{2}$  & 5.89  & 3.71  & 4.51  & 3.30 \tabularnewline
\hline 
total $\chi^{2}$  & 3838.95  & 3835.63  & 3841.52  & 3838.00 \tabularnewline
\hline 
\end{tabular}\caption{Constraints ($68\%$) and best-fit values in parentheses on the main
parameters based on CMB$+$BAO$+$Pantheon$+$R19$+$BBN. The values
of $\chi^{2}$ in the lower rows are for the best-fit points in each
model.}
\label{Tab:Bestfit} 
\end{table}

\subsection{Discussions}

\begin{figure}
\includegraphics[width=12cm]{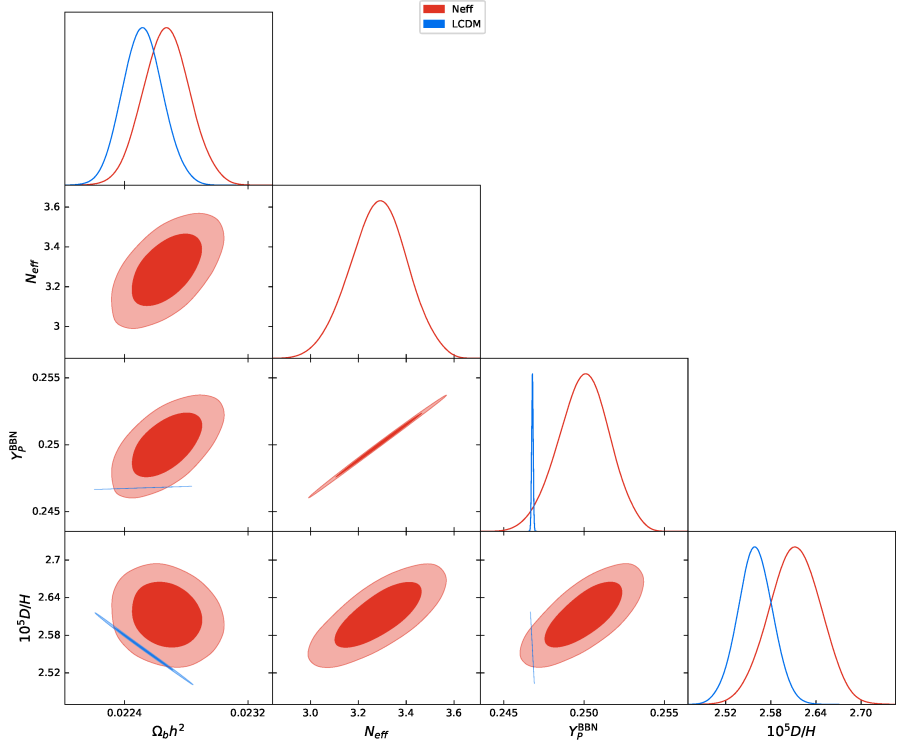} \caption{Posterior and its BBN dependence in the $N_{\mathrm{eff}}$ model.}
\label{Fig:BBN:Neff} 
\end{figure}

\begin{figure}
\includegraphics[width=12cm]{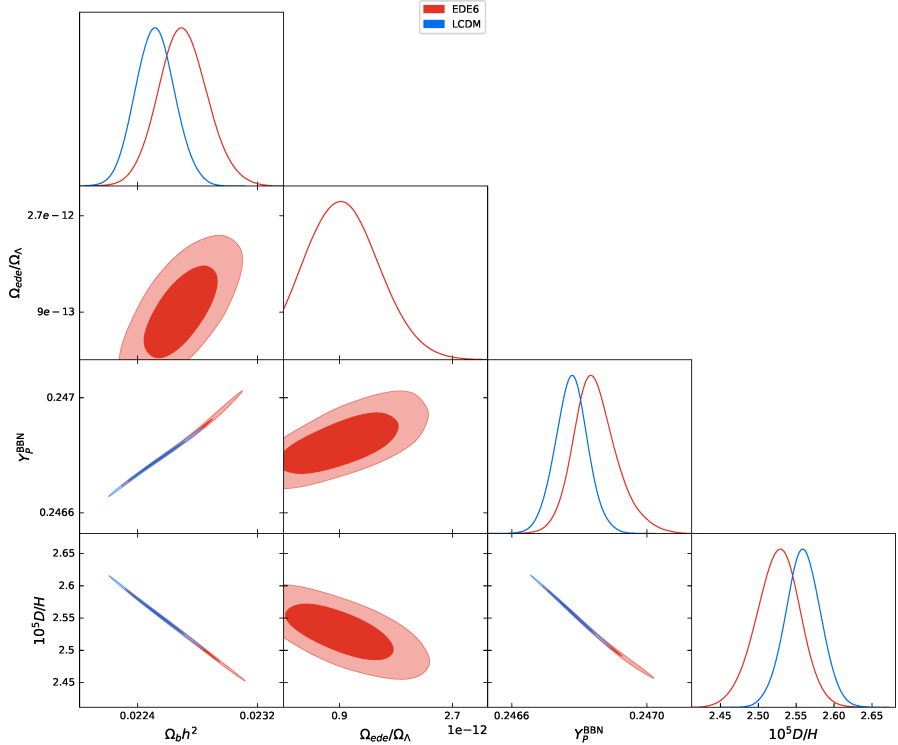} \caption{Posterior and its BBN dependence in the EDE$6$ model.}
\label{Fig:BBN:EDE} 
\end{figure}

As is well known, an increase of $N_{\mathrm{eff}}$ affects the fit
with the observations of light elements, because it contributes the
cosmic expansion at the BBN epoch and alters the $p/n$ ratio. This
leads to an increase in both the helium mass fraction $Y_{P}$ and
deuterium abundance D/H. By increasing $N_{\mathrm{eff}}$, the CMB
fit simultaneously indicates a larger $\Omega_{b}h^{2}$ which reduces
D/H. In total, the enhancement of D/H is suppressed. As a result,
the $\chi^{2}$ of $Y_{P}$ observations ($\chi_{\mathrm{Aver}}^{2}$)
increases, while the $\chi^{2}$ of D/H observations ($\chi_{\mathrm{Cooke}}^{2}$)
increases a little. In fact, the value of $\chi_{\mathrm{Cooke}}^{2}$
of the $N_{\mathrm{eff}}$ model is smaller than that in the $\Lambda$CDM
model. This can be seen in Fig.~\ref{Fig:BBN:Neff} and Table~\ref{Tab:Bestfit}.
On the other hand, in the EDE scenarios, increasing $\Omega_{b}h^{2}$
to adjust the CMB fit reduces the D/H abundance significantly. Thus,
$\chi_{\mathrm{Aver}}^{2}$ increases a little, while $\chi_{\mathrm{Cooke}}^{2}$
increases. This can be seen in Fig.~\ref{Fig:BBN:EDE} and Table~\ref{Tab:Bestfit}.
The tradeoff relation between the fit to the helium mass fraction
$Y_{P}$ and deuterium abundance D/H can be seen more clearly in Table~\ref{Tab:Bestfit},
where we compare it with the coexistence model.

%[htbp]

%[htbp]

The constraint on $N_{\mathrm{eff}}$ from BBN and BAO data only without
including CMB or SH0SE data has been derived as $N_{\mathrm{eff}}=2.88\pm0.16$,
which is less than $3$~\cite{Cyburt:2015mya}. As is well known,
a larger $N_{\mathrm{eff}}$ is disfavored by BBN. This is consistent
with the results in Table~\ref{Tab:Bestfit}. What we additionally
find is that the EDE models without $\Delta N_{\mathrm{eff}}$ are
also limited by BBN, because they predicts too little D/H abundance
by too large $\Omega_{b}h^{2}$.

\begin{figure}[htbp]
\includegraphics[width=13cm]{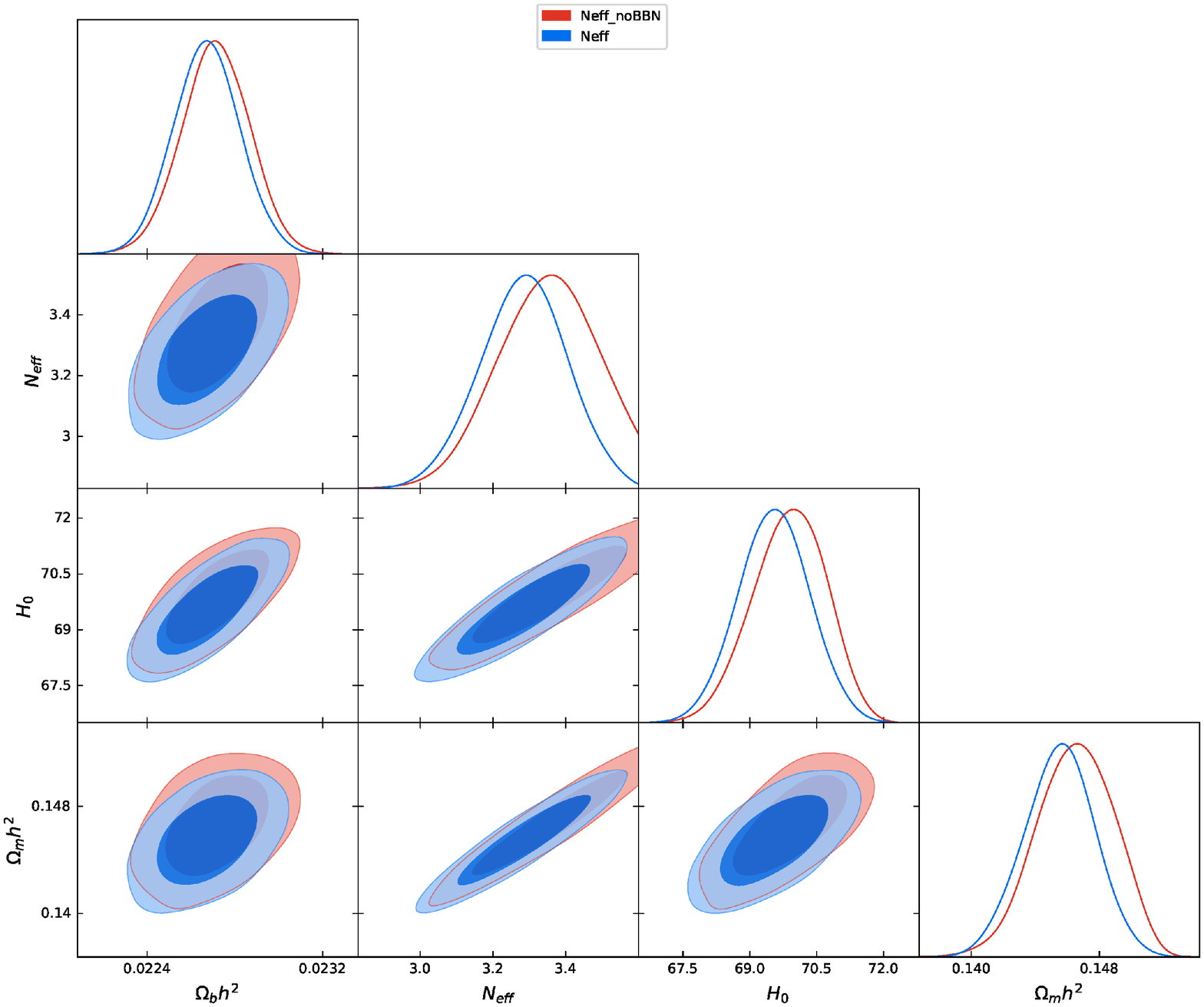} %\caption{The posterior and its BBN dependence in a $N_\mathrm{eff}$ model.}
%\label{Fig:BBN:Neff}
%\end{figure}
%%
%\begin{figure}[htbp]
\includegraphics[width=13cm]{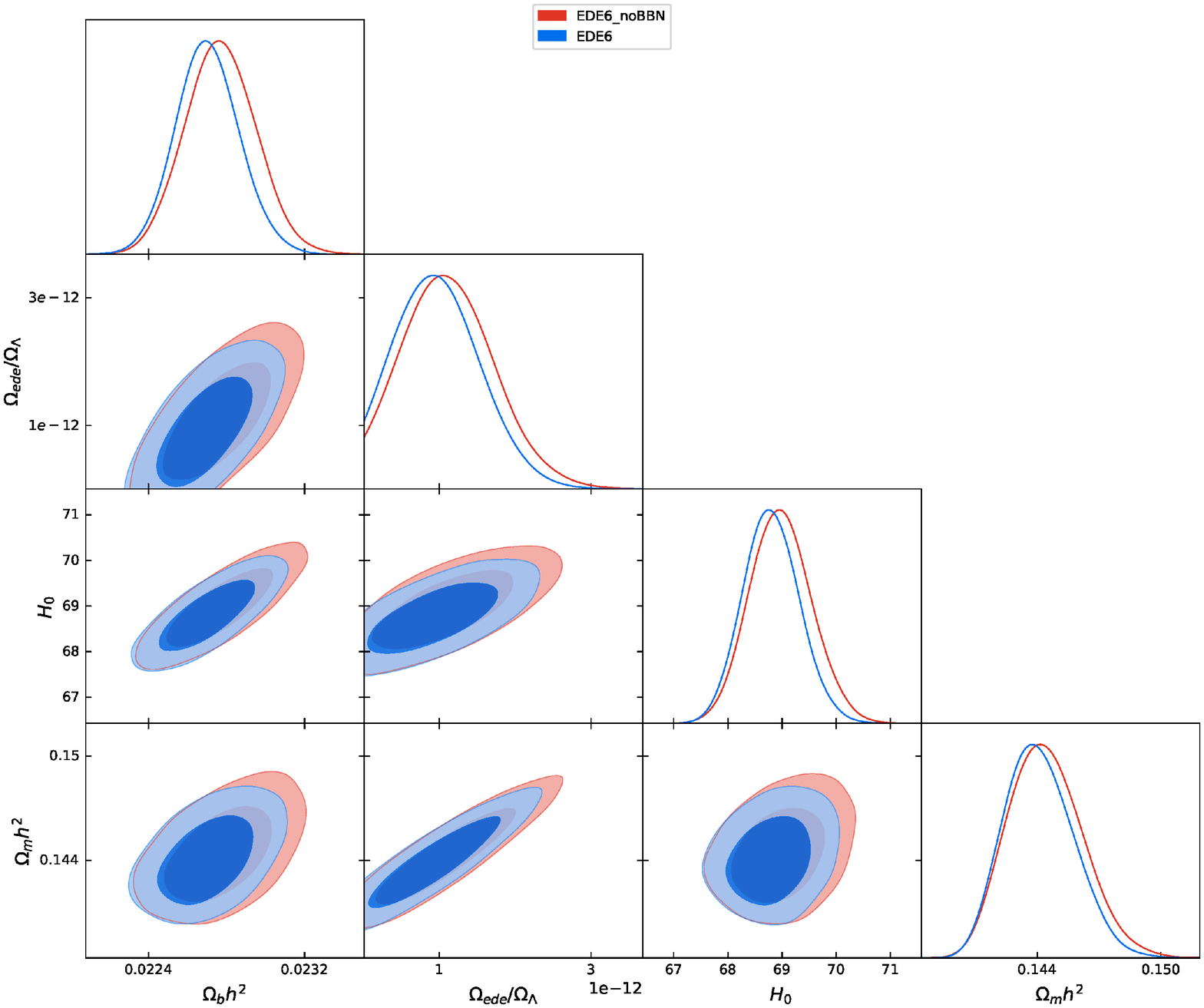} %\caption{The posterior and its BBN dependence in a EDE$6$ model.}
%\label{Fig:BBN:EDE}
%\end{figure}
\caption{Comparison of constraints based on data sets with and without BBN
for the $N_{\mathrm{eff}}$ (upper) and EDE6 (lower) models.}
\label{Fig:with-out-BBN} 
\end{figure}

In the literature on the new physics interpretation of the Hubble
tension, data sets have not included BBN data. We have derived constraints
from the data sets with and without BBN data for each model. As can
be expected, data sets without BBN indicate larger values of $H_{0}$
and $N_{\mathrm{eff}}$. %The magnitude of differences are shown in Tab.~\ref{Tab:with-out-BBN}.
%\begin{table}[htbp]
%\begin{tabular}{|c||c|c|c|c|c|}
%\hline 
%Model & EDE$6$ & EDE$6$+$N_\mathrm{eff}$ & $\Lambda$CDM & $N_\mathrm{eff}$    \tabularnewline
%\hline
%\hline 
%$H_0$ [km/s/Mpc] without BBN & $68.97 \pm 0.57$ & $70.03\pm 0.87$ & $68.22 \pm 0.41$ & $69.89^{+0.89}_{-0.76}$ \tabularnewline
%\hline 
%$N_{\textrm{eff}}$ without BBN & $3.046$ & $3.27^{+0.16}_{-0.14}$ & $3.046$ & $3.34^{+0.15}%{-0.12}$ \tabularnewline
%\hline 
%\hline 
%$H_0$ [km/s/Mpc] with BBN & $68.80 \pm 0.52$ & $69.73\pm 0.82 $ & $68.19\pm 0.39$  & $69.54\pm %0.80$   \tabularnewline
%\hline 
%$N_{\textrm{eff}}$ with BBN & 3.046 & $3.23 \pm 0.13$ & 3.046 & $3.28\pm 0.12$  \tabularnewline
%\hline 
%\end{tabular}
%\caption{Comparison of constraints based on data sets with and without BBN.}
%\label{Tab:with-out-BBN}
%\end{table}
The magnitudes of the differences are shown in Fig.~\ref{Fig:with-out-BBN}.

%An interesting implication is that the preferred scalar spectral index $n_{s}$
%is increased as large as $0.98$, as has been reported, for example, in Ref.~\cite{Braglia:2020bym} as well.  
%The Planck result~\cite{Aghanim:2018eyx} without including local measurements of 
%the Hubble constant indicates $n_{s}=0.9665\pm0.0038\;$(68\%,TT,TE,EE+lowE+lensing+BAO) 
%and disfavors spontaneously broken supersymmetric inflation~\cite{Dvali:1994ms} 
%predicting $n_s \simeq 0.98$.
%On the other hand, cosmology with $N_\mathrm{eff}$ or the EDE seems to be compatible with $n_{s}=0.98$.

\section{Summary}

A shorter sound horizon scale at the recombination epoch, arising
from introducing extra energy components such as extra radiation or
EDE is a simple approach to resolving the so-called Hubble tension.
However, then the compatibility with successful BBN would be a concern,
because the extra radiation may contribute to the cosmic expansion
or the inferred baryon asymmetry would be different from that in the
$\Lambda$CDM. We have compared the EDE models, $N_{\mathrm{eff}}$
model, and a coexistence model, paying attention to the fit to BBN.
In fact, the EDE models are also subject to the BBN constraints by
increasing the order-unity $\chi^{2}$ as in the $N_{\mathrm{eff}}$
model. Our main results are summarized in Fig.~\ref{Fig:Comparison}
and Table~\ref{Tab:Bestfit}. By comparing the posteriors based on
the CMB$+$BAO$+$Pantheon$+$R19 combined data, both the $N_{\mathrm{eff}}$
and a coexistence models indicate the largest $H_{0}$ between all
of the models studied. $H_{0}$ can be as large as $70.5$ km/s/Mpc
within $1\sigma$ only for the coexistence model. The goodness of
the fits for the models in terms of $\chi^{2}$ are also listed in
Table~\ref{Tab:Bestfit}. The fitting is good in the order of the
$N_{\mathrm{eff}}+$EDE$6$, $N_{\mathrm{eff}}$, and EDE$6$ models
and the $\Lambda$CDM. The difference of the best-fit $H_{0}$ values
between the $N_{\mathrm{eff}}+$EDE$6$ and $N_{\mathrm{eff}}$ models
is tiny, while the $\chi^{2}$ difference is about $2.4$. The extra
radiation seems to be more effective at causing a large $H_{0}$ than
the EDE model. Thus, the $N_{\mathrm{eff}}$ model is a much simpler
and better model than the EDE models.

We also examined the data sets dependence, whether we include BBN
or not. The difference on $N_{\mathrm{eff}}$ in the $N_{\mathrm{eff}}$
model is only about $0.06$ in its mean value, however, the including
errors indicate 
\begin{align}
 & 3.22<N_{\mathrm{eff}}<3.49\,(68\%)\quad\mathrm{for\quad(CMB+BAO+Pantheon+R19)},\\
 & 3.16<N_{\mathrm{eff}}<3.40\,(68\%)\quad\mathrm{for\quad(CMB+BAO+Pantheon+R19+BBN)},
\end{align}
and there is almost $0.1$ difference in the upper. By comparing this
with Eq.~(\ref{Eq:PlanckNeff}), we can see the impact of the R$19$
data compared to the R$18$ data. For EDE models, if we include the
BBN data, a smaller $H_{0}$ and smaller EDE energy density are preferred.

%======================================%
%<<<<<<<<<< ACKNOWLEDGMENTS >>>>>>>>>>>%
%======================================%

\section*{Acknowledgments}

We would like to thank T.~Sekiguchi for kind correspondences concerning
the use of CosmoMC. This work is supported in part by the Japan Society
for the Promotion of Science (JSPS) KAKENHI Grants Nos.~19K03860,~19H05091,
and~19K03865 (O.S.).

%%%%%%%%%%%%%%%%%%%%%%%%%%%%
%   Planck
%%%%%%%%%%%%%%%%%%%%%%%%%%%%
%\cite{Aghanim:2018eyx}

\end{document}